# Strain distribution and thermal strain relaxation in MOVPE grown hBN films on sapphire substrates

Kousik Bera,[1] D. Chugh,[2] Atanu Patra,[3] H. Hoe Tan,[2] C. Jagadish[2,4] Anushree Roy[3]

[1]*Center for Nanoscience and Technology, Indian Institute of Technology Kharagpur. Pin 721302. India.*

[2]*epartment of Electronic Materials Engineering, Research School of Physics and Engineering, The Australian National University, Canberra, Australian Capital Territory, Australia*

[3]*Department of Physics, Indian Institute of Technology Kharagpur. Pin 721302. India.*

[4]*Australian National Fabrication Facility, Research School of Physics and Engineering, The Australian National University, Canberra, ACT 2601, Australia*

## Abstract

Recently, hexagonal boron nitride (hBN) layers have generated a lot of interest as ideal substrates for 2D stacked devices. Sapphire-supported thin hBN films of different thicknesses are grown using metalorganic vapour phase epitaxy technique by following a flow modulation scheme. Though these films of relatively large size are potential candidates to be employed in designing real devices, they exhibit wrinkling. The formation of wrinkles is a key signature of strain distribution in a film. Raman imaging has been utilized to study the residual strain distribution in these wrinkled hBN films. An increase in the overall compressive strain in the films with an increase in the layer thickness has been observed. To find whether the residual lattice strain in the films can be removed by a thermal treatment, temperature dependent Raman measurements of these films are carried out. The study demonstrates that the thermal rate of strain evolution is higher in the films of lower thickness than in the thicker films. This observation further provides a possible explanation for the variation of strain in the as-grown films. An empirical relation has been proposed for estimating the residual strain from the morphology of the films. We have also shown that the residual strain can be partially released by the delamination of the films.

## Introduction

Recently, thin layers of hexagonal boron nitride (hBN) have attracted remarkable attention for their unique physical properties [1]. With hexagonal rings of alternating boron and nitrogen atoms in the $sp^2$ network, they are very similar to graphene. Atomically smooth surface [2], absence of dangling bond [3], superior thermal [4,5] and chemical stability [5] make them ideal candidates as insulating substrates in 2D electronics and for growing multilayer heterostructures. For the fabrication of graphene-based devices, a thin layer of hBN serves not only as an insulating substrate but also as an encapsulating layer [2,6–8]. hBN supported graphene layers have significantly better carrier mobility than the devices on $SiO_2$ [2]. In spite of being an indirect band gap material, hBN can emit deep UV wavelength, originated from sub band gap defect states [9]. The use of 2D hBN as a buffer layer with other 2D materials may lead to development of a wide range of advanced electronic and optoelectronic devices.

While studying a 2D system, most of the reports in the literature deal with hBN layers obtained from mechanical exfoliation of bulk crystals. These films find limitations in many applications due to the restricted availability of a large surface area. In recent studies [10,11] the possibility of large scale growth of hBN films using metal organic vapour phase epitaxy (MOVPE) technique has been discussed. The growth parameters are varied to tune the characteristics of the films. Though, these films, grown on a relatively large area, are potential candidates as dielectric layers for 2D material based field effect device fabrication, wrinkling has been observed in as-grown layers. Wrinkling indicates the possibility of a residual strain distribution in the films [12, 13].

Raman spectroscopy is an indispensable technique to probe the lattice strain [14-18] in 2D materials. In general, the dynamics of phonons is extremely rich for these systems. In case of graphene, G (doubly degenerate $E_{2g}^{high}$ phonon mode) and 2D (originating from an intervalley fourth order process, in which two phonons with opposite wavevectors participate) peaks are the characteristic Raman modes, which reveal many of the essential physics related to the system [19,20]. The gliding motion of the hexagonal planes is responsible for the low frequency $E_{2g}^{low}$ interlayer share mode in graphene. The honeycomb structure of hBN is related to graphene, except that boron and nitrogen follow intra-plane AA′ stacking (AA in graphene). Similar to graphene, the Raman active inplane mode ($E_{2g}^{high}$) and interlayer shear mode ($E_{2g}^{low}$) of hBN reveal detail characteristics of the system. Interestingly, these two modes behave differently under external perturbation (stress, thermal treatment) [21,22]. For example, the red-shift of the $E_{2g}^{high}$ phonon frequency with temperature can only be explained by taking into account the negative thermal expansion of the material along with anharmonic phonon decay in the films [21]. In contrast, the softening of the $E_{2g}^{low}$ phonon mode with the increase in temperature mainly carries the signature of the increase in interlayer distance. The doubly degenerate $E_{2g}^{high}$ mode splits under uniaxial tensile strain [22]. Under biaxial strain, this mode shifts with a remarkably high rate for thin hBN layers [22]. Recently, Stenger *et al.* [23] have shown that the Raman shift of the interlayer $E_{2g}^{low}$ mode characterizes the thickness of hBN films.

In this article, we have studied the strain distribution in sapphire-supported hBN films of different thicknesses, grown by FM-MOVPE technique. The positive and negative thermal expansion coefficients of sapphire and hBN lead to the formation wrinkles in the films, which are characterized by atomic force microscopy (AFM). The residual non-uniform strain distributions in the films are investigated by Raman mapping. It is to be noted that Raman mapping is an invaluable tool to study the spatial characteristics of these anisotropic films, compared to what can be achieved by studying spectra at arbitrarily chosen points on the films. We have observed an overall compressive strain in the films from the map of the Raman shift of the above discussed $E_{2g}^{high}$ mode of hBN. We have noted an increase in the residual strain with an increase in the film thickness. Temperature dependent Raman measurements are carried out to examine whether the residual strain can be eliminated by a thermal treatment and to explain the observed variation of the same with the thickness of the films. We have also estimated the temperature range to scan for obtaining the nearly strain-free sapphire-supported hBN films of different thicknesses. An empirical relation has been proposed for estimating the residual strain from the morphology of the films. Furthermore, it has been demonstrated that the compressive strain in the films can be partially removed by delamination.

**Experimental details**

hBN films were deposited on 2″×2″ sapphire substrates using MOVPE growth technique. The growth was carried out in Aixtron MOVPE close-coupled showerhead (CCS) reactor, where triethylboron (TEB) and ammonia ($NH_3$) were used as boron and nitrogen precursors, respectively. The flow modulation (FM) growth mechanism was followed to grow smooth hBN film. Under this scheme, TEB and $NH_3$ flow were introduced in the reactor as short alternating pulses for 1 sec and 3 sec., respectively. The substrate temperature was kept fixed at 1350 ˚C during the growth process. The thicknesses of the films were controlled by varying the growth duration. The detail growth procedure is available in Ref. [10].

Experiments were also carried out on hBN films transferred on $SiO_2$/Si substrates. In the presence of de-ionized water the MOVPE grown hBN films self-delaminate from the substrate. After complete detachment, the films were transferred on the $SiO_2$/Si matrix. The details of the growth and film transfer processes are available in detail in Ref [10].

A tapping mode AFM (SCANASYSTAIR, MultiMode 8, Bruker) was used to observe the surface morphology and to measure the film thickness. For micro-Raman spectroscopy, a triple monochromator spectrometer (model T64000, make Horiba, France) was used in the subtractive mode. The unit is equipped with a Peltier cooled CCD detector (model Synapse, Make Horiba, France), a water-cooled Argon-Krypton ion laser (Model Innova 70C Spectrum, Coherent, USA) and an optical microscope (BX41, Olympus, Japan). To collect the backscattered radiation, 100× (numerical aperture 0.90) and 50×L (numerical aperture 0.50) objective lenses were used for room temperature and temperature dependent measurements, respectively. The integration time to record the spectra was varied from a few minutes to hours with a decrease in the film thickness. Here we would like to mention that with T64000 spectrometer, the Raman intensity of the thinnest hBN film, probed in this study, was extremely poor for analysis. Thus, the spectra of this film were also recorded using a single monochromator spectrometer (model iHR550, make Horiba, France) for better signal to noise ratio. Point by point Raman mapping on films was carried out using a computer controlled motorized XY stage. We mapped 1.98 μm×0.88 μm area of all films with a step size of 0.220 μm. High temperature sample stage (model TS1500, Make Linkam Scientific, UK) was used to carry out the temperature-dependent Raman measurements. Before recording the Raman spectrum at each temperature, we waited for 10-20 mins after the temperature was stabilised.

All experiments were carried out using 514 nm as the excitation wavelength. The band gap (>5.5 eV) [9] of hBN is much higher than the available excitation energies in visible Raman spectroscopy. Therefore, the inelastically scattered Raman signal is very weak from the films. To enhance the intensity of the Raman signal, incident laser power cannot be increased beyond a certain limit as that may lead to sample heating [23]. To optimize the laser power, required to obtain reasonably good signal to noise ratio without laser heating of the samples, we recorded Raman spectra at various incident powers (not shown). After conducting this test, the individual spectra of the films were recorded with 2.8 mW power on the samples in our experiments. For Raman mapping, the samples were exposed to radiation for a much longer time. For example, the total experimental duration was 6 hrs for a 20 nm thick hBN film for Raman mapping of the area 1.98 μm × 0.88 μm with a step size of 0.220 μm. Thus, the incident laser power was further reduced while carrying out Raman mapping. After completion of the scan of the desired area, spectra were recorded again in the first few points of the mapped area to ensure that we get similar spectra as obtained at the beginning of the scan.

## Results

### A. Wrinkles in hBN films on sapphire substrates

While cooling the reactor chamber, wrinkles develop in the FM-MOVPE grown hBN films on sapphire substrates. Figure 1(a)-(c) exhibit characteristic AFM images of some of the films, under study, grown for different time duration. The colour scale in the right of each panel marks the nanoscale topographical information of the surface of the films. The formation of wrinkles throughout each film is observed. From the height line profiles (not shown) we have measured the thickness ($h$) of the films. In this article we study the films of thicknesses, 2±1 nm, 5±2 nm, 10±2 nm, 20±3 nm and 40±3 nm.

To describe the physics of wrinkle formation, two parameters, wrinkle wavelength ($\lambda$) and wrinkle amplitude ($A$) [12,13], are often used (see the inset of Figure 1(d)). It is to be noted in Figure 1(a)-(c) that the average values of both $\lambda$ and $A$ of wrinkles are different in the films of different thicknesses. The variations of the mean values of $\lambda$ and $A$ with the thickness ($h$) of the films, as measured from the AFM images, are shown by orange and violet symbols in Figure 1(d). The error bars in $\lambda$ are the standard deviations of the values of the parameter obtained for ~80 wrinkled domains for each film. The orange solid line is a guide to the eyes to follow the $\lambda$ vs. $h$ plot. The unavoidable parasitic reactions during the growth process form debris (mostly hBN nanoparticles) on the films (see Ref. 10 for details), which appear as bright spots in the AFM images. Thus, it is difficult to comment on the trend followed by the data points in the *A* vs. *h* plot.

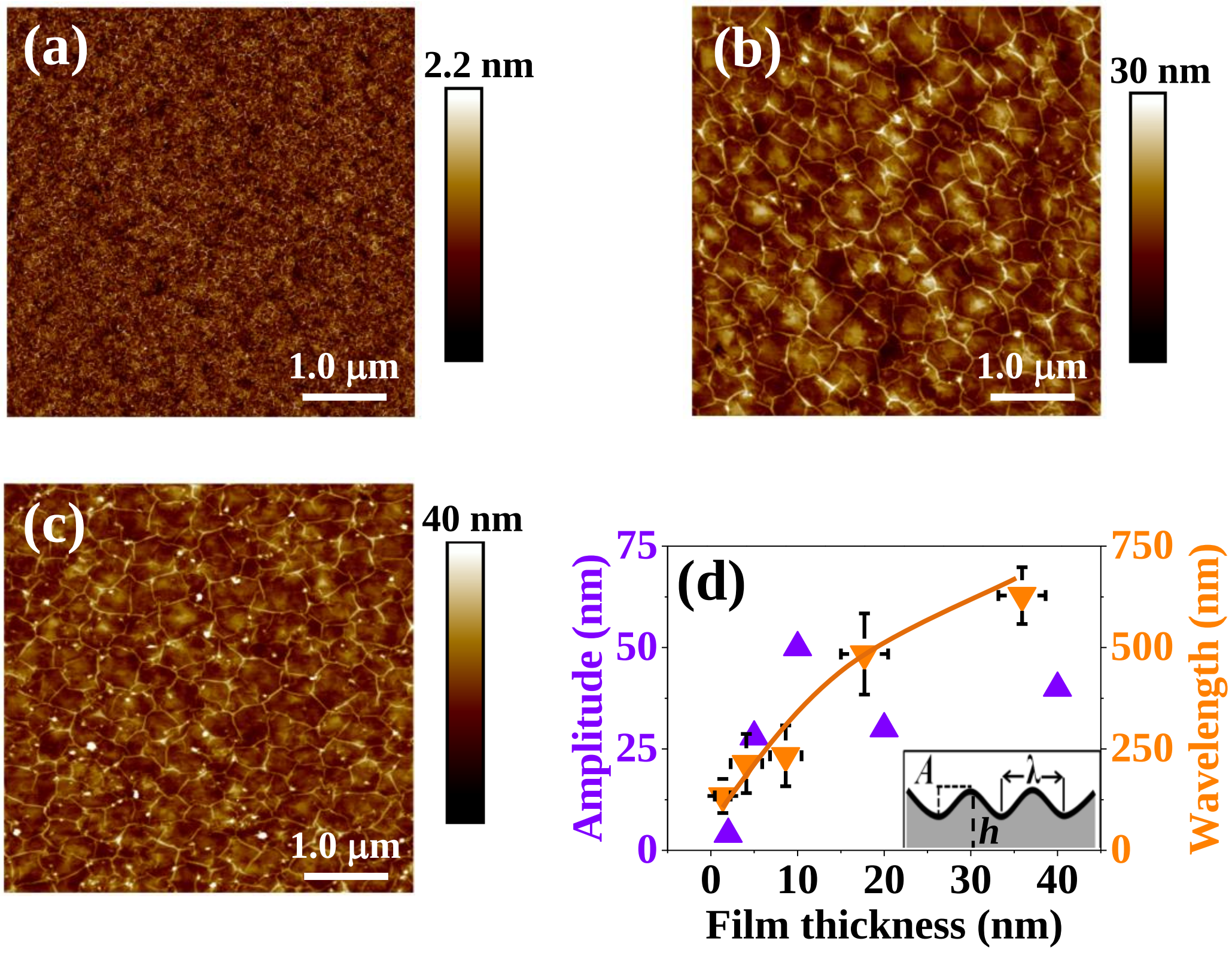

**Figure 1. Surface morphology of (a) 2 nm, (b) 20 nm and (c) 40 nm thick hBN films on sapphire substrates at room temperature. (d) The variation of average value of the wavelength, λ, (right scale) and amplitude, *A*, (left scale) with the film thickness, *h*. The solid line is a guide to the eyes to follow the change in *λ* with *h*. Inset of (d) schematically defines the wrinkle wavelength *λ* and amplitude *A* and thickness (*h*).**

### B. Residual strain in the films

Wrinkling may leave a residual strain in the film. As mentioned earlier, the phonon dynamics of a 2D material is highly sensitive to lattice strain and crystalline quality of the films, which can be probed by Raman spectroscopic measurements. The normalized Raman spectra of the films over the spectral range between 1250 $cm^{-1}$ and 1500 $cm^{-1}$ are shown in Figure 2. The peak of each spectrum corresponds to the $E_{2g}^{high}$ vibrational mode of hBN, discussed earlier.

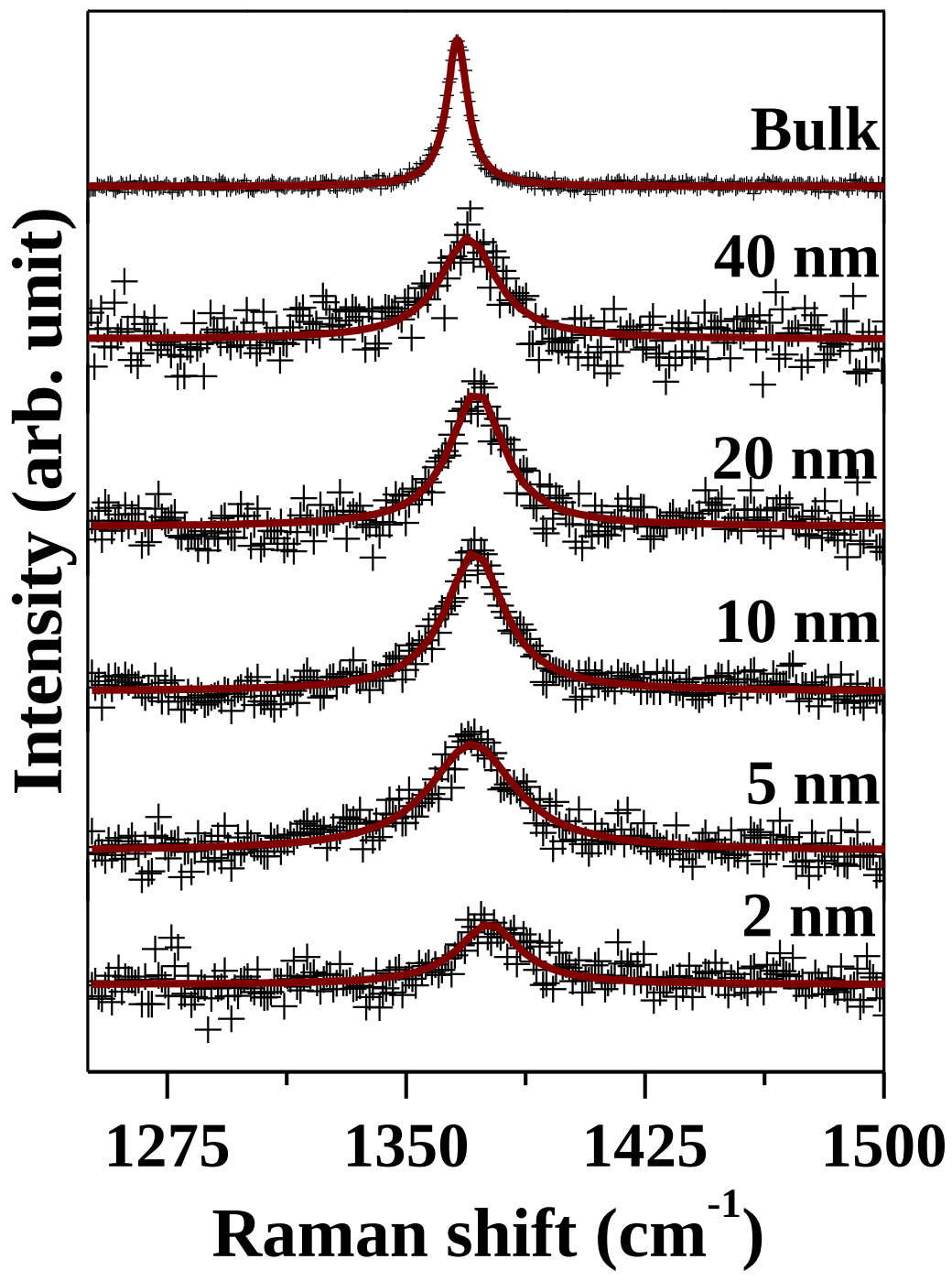

**Figure 2. Raman spectra of hBN films of different thicknesses. The top most panel shows the Raman spectrum of bulk hBN. In each panel, the net fitted spectrum, obtained by fitting the data points using a Lorentzian function is shown by the solid line.**

Each spectrum is fitted with a Lorentzian function and shown by the solid red line in Figure 2. The Raman spectrum of bulk hBN is shown in the top for reference. Here we would like to mention that we did not observe the $E_{2g}^{low}$ mode at 52 $cm^{-1}$, most probably, due to its low intensity in our non-resonant and low power experimental conditions.

The spatial non-uniform and wrinkled surface topography of the films, as shown in Figure 1, indicates that the structural characteristics may not be uniform over the film area. Thus, the above single point Raman measurements (shown in Figure 2) may not reveal the true properties of the films. This motivated us to carry out Raman mapping, over a reasonably large area encompassing minimum two wrinkled domains. In this imaging technique, spectra are collected at each point of a raster-scanned sample. False colour images, which are based on the spectral parameters of the characteristic Raman mode of the films, are then generated.

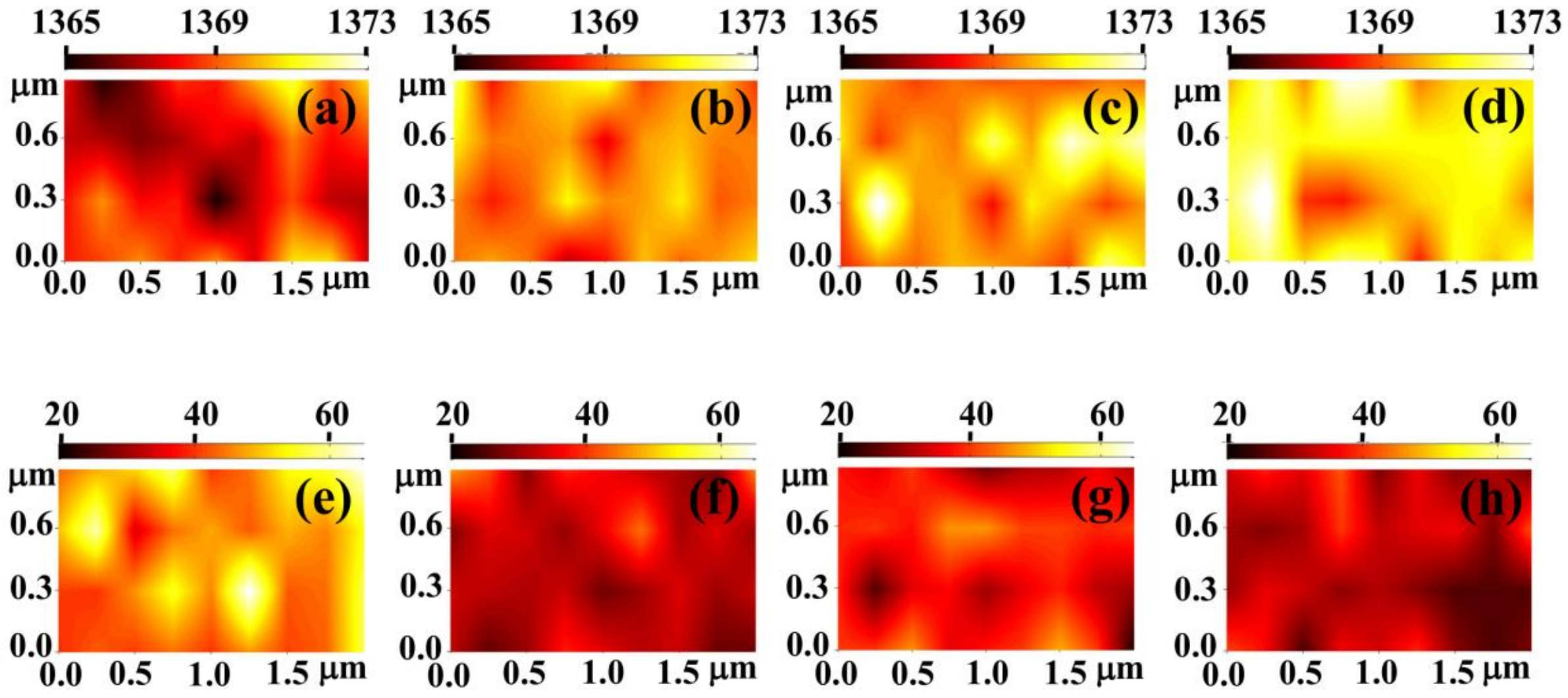

**Figure 3. Mapping of the Raman shift of the** $E_{2g}^{high}$ **mode for (a) 5±2 nm, (b) 10±2 nm, (c) 20±3 nm and (d) 40±3 nm hBN films. The corresponding FWHM of the** $E_{2g}^{high}$ **phonon mode are shown in (e)-(h), respectively. The scale bars are in** $cm^{-1}$**.**

Figure 3(a)-(d) present characteristic map of Raman shift of the $E_{2g}^{high}$ vibrational mode of hBN films of thicknesses 5±2 nm, 10±2 nm, 20±3 nm and 40±3 nm, respectively. The maps of full width at half-maximum (FWHM) of the same mode of these films are shown in Figure 3(e)-(h). The colour bar on the top of each panel exhibits the scale (in $cm^{-1}$) for Raman wavenumber (a-d) and peak width (e-h) of the $E_{2g}^{high}$ Raman mode of hBN in these maps. While choosing the area of the film for mapping, we purposely avoided the debris (appeared as white regions in AFM images in Fig. 1). Even then Raman spectral parameters exhibit non-uniformity over the scanned area. For a better comparison of the change in these parameters with the thickness of the films, the scale bars for Raman shift and width are kept same for all images. The saturation in colour in some of the images (say for Figure 3(c), (d) and (e)) indicates that the ranges of the measured parameters are greater than the range used in the scale bar.

We also find that the distributions of Raman wavenumber and width vary with film thickness. For example, while for the 5 nm hBN film, the wavenumber over the given scanned area vary from 1360 $cm^{-1}$ to 1372 $cm^{-1}$ (over 12 $cm^{-1}$), for the 40 nm film the range is between 1368 $cm^{-1}$ and 1373 $cm^{-1}$ (over 5 $cm^{-1}$). The corresponding distribution of the FWHM are from 34 $cm^{-1}$ to 104 $cm^{-1}$ for the 5 nm film (over 70 $cm^{-1}$) and from 24 $cm^{-1}$ to 42 $cm^{-1}$ (over 18 $cm^{-1}$) for the 40 nm film. Wrinkling leaves a non-uniform strain distribution within the wrinkled domain, which can vary with the thickness of the films [24]. It is also to be noted that we have scanned a fixed area for all films with the laser spot size of nearly 1 μm. As the wrinkled domain size varies with the film thickness (shown in Figure 1), we have probed different number of domains for each of these films in our experiment. Thus, it is non-trivial to relate the observed variation in Raman shift and width to the domain characteristics of these films.

Figure 4(a) and (b) plot the variation of the mean value (symbols) and standard deviation (vertical error bars) of the Raman shift and FWHM with the film thickness over the scanned area. The measured Raman wavenumber (1367 $cm^{-1}$) and FWHM (9 $cm^{-1}$) of the $E_{2g}^{high}$ phonon mode of the bulk hBN are shown by dashed lines in 4(a) and (b), respectively. We observe a blue shift in Raman wavenumber of the films from its bulk value. It is also interesting to note that though the standard deviation of some of the data points are large, there is an increasing trend in Raman shift with an increase in the thickness of the films. In contrast, the mean FWHM decreases with an increase in film thickness, as shown in Figure 4(b).

To probe the origin of the observed characteristics of the residual strain in the films in Figure 3 and also to find whether this strain can be eliminated by varying the temperature of the films, we performed temperature dependent Raman measurements. The crystal structure depends on temperature and the phonon dynamics of 2D films strongly gets affected by the temperature of the system. In this study, Raman spectra were collected from a single point at different temperatures. In the beginning of the measurements for each film, we have carefully chosen the point on the film, for which the Raman shift was very close to the mean value obtained in Figure 4(a).

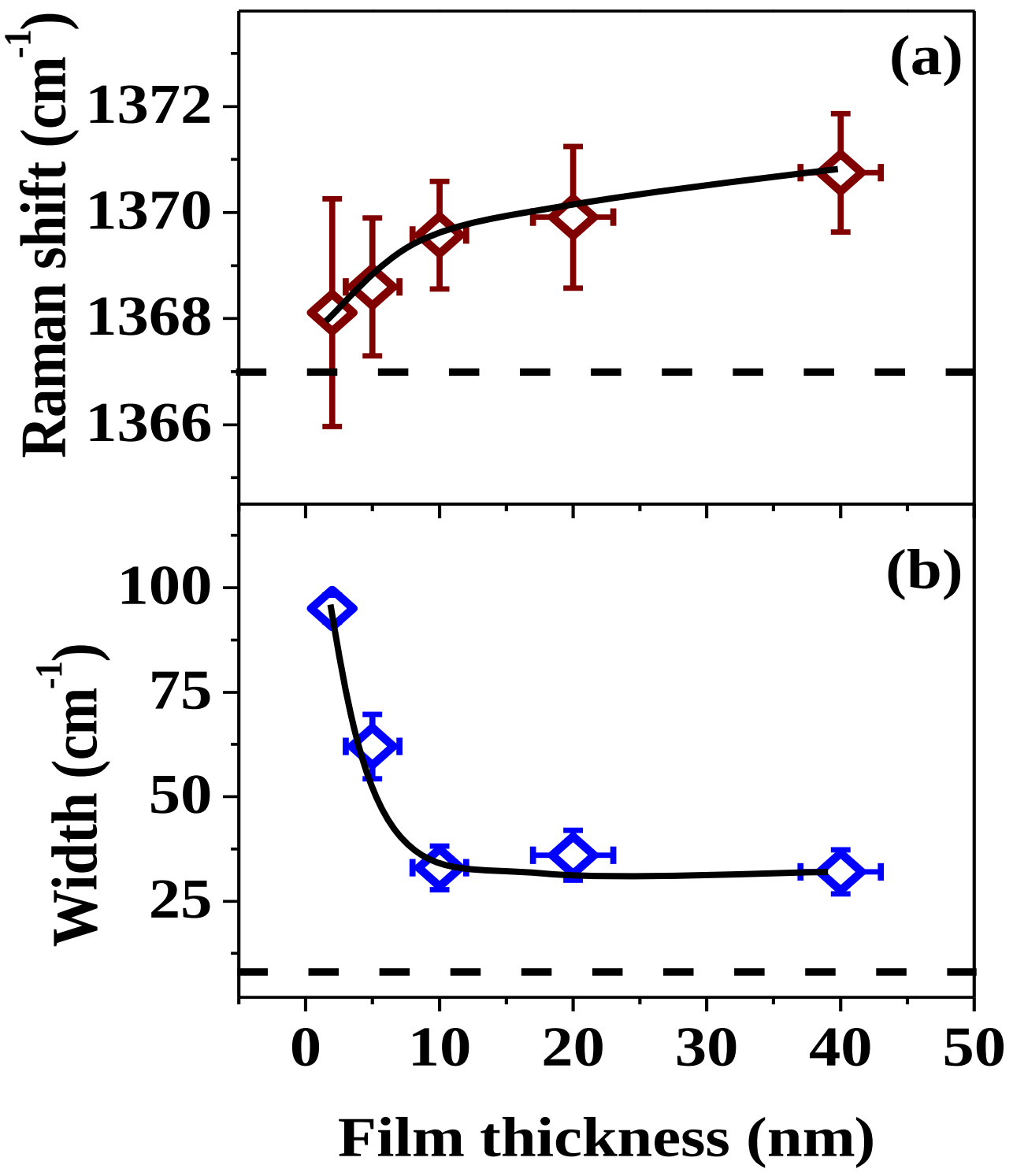

**Figure 4. The variation of the mean value of (a) Raman shift and (b) FWHM of the $E_{2g}^{high}$ mode with the thickness of the film, as obtained by analysing each spectrum of the mapped area. The vertical error bar in each panel is the standard deviation of all values obtained from the image frame. The dashed line in (a) marks the Raman shift of bulk hBN. The same in (b) marks the FWHM of bulk hBN. The solid line in each panel is a guide to the eyes.**

Raman spectra of the films at these chosen points and bulk hBN were then recorded at different temperatures over the spectral window between 1250 $cm^{-1}$ and 1500 $cm^{-1}$. Each spectrum was fitted with a Lorentzian function, keeping the peak position, width and intensity as free fitting parameters. The variation in Raman shift of the $E_{2g}^{high}$ phonon mode with temperature for bulk hBN is shown in Figure 5 (a) and the same for films of different thicknesses are shown by symbols in Figure 5 (b)-(f). The error bars to the data points are the standard deviation of the Raman shift obtained from the fitting procedure.  It is to be noted that the rate of down shift of the Raman shift with temperature varies with the thickness of the film. We have carried out high temperature Raman mapping of 2 nm and 40 nm films. It is indeed found that the Raman shift vs. temperature plots undergo only a vertical shift, however, the nature remains same.

Here we would like to mention that we have observed a strong rise in luminescence background for the films with increasing temperature. Sapphire exhibits thermoluminescence peak in the visible range [25]. The sub-band gap defect states of hBN can also give rise to visible luminescence from the films [26]. Beyond a certain temperature, the rising background completely masked the Raman signal of the films.

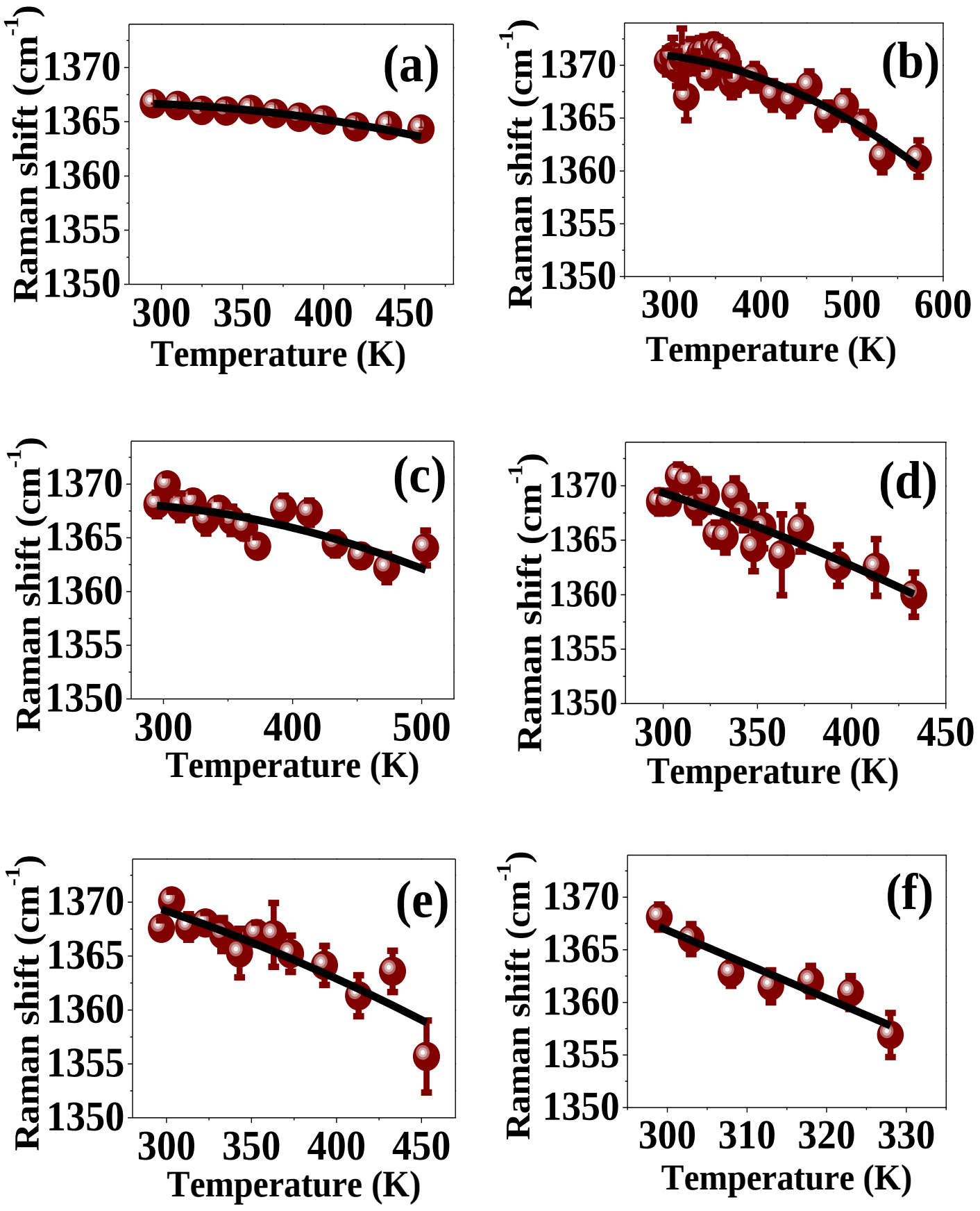

**Figure 5. Temperature dependent Raman shift of (a) bulk hBN film and films of thickness (b) 40 nm (c) 20 nm (d) 10 nm (e) 5 nm and (f)  2 nm.**

### C. Study on delaminated hBN films

To investigate the residual strain distribution in the delaminated films, we have transferred films on $SiO_2$/Si. Figure 6(a) shows a characteristic AFM image of the delaminated film of thickness 20 nm. By comparing the AFM image of the substrate supported same film in Figure 1(b), we note that the delaminated film is free from wrinkles. A similar change in surface morphology has been observed in AFM images of 40 nm thick hBN film as well, which undergo self-delamination from sapphire in presence of DI water. Raman mapping was carried out on the transferred 20 nm thick hBN film. Map of Raman shift of the $E_{2g}^{high}$ phonon mode of the transferred hBN film of thickness 20 nm are shown in Figure 6(b). By comparing the same of the substrate-supported film in Figure 3(c), we find a significant decrease in the distribution of Raman shift in the transferred film.

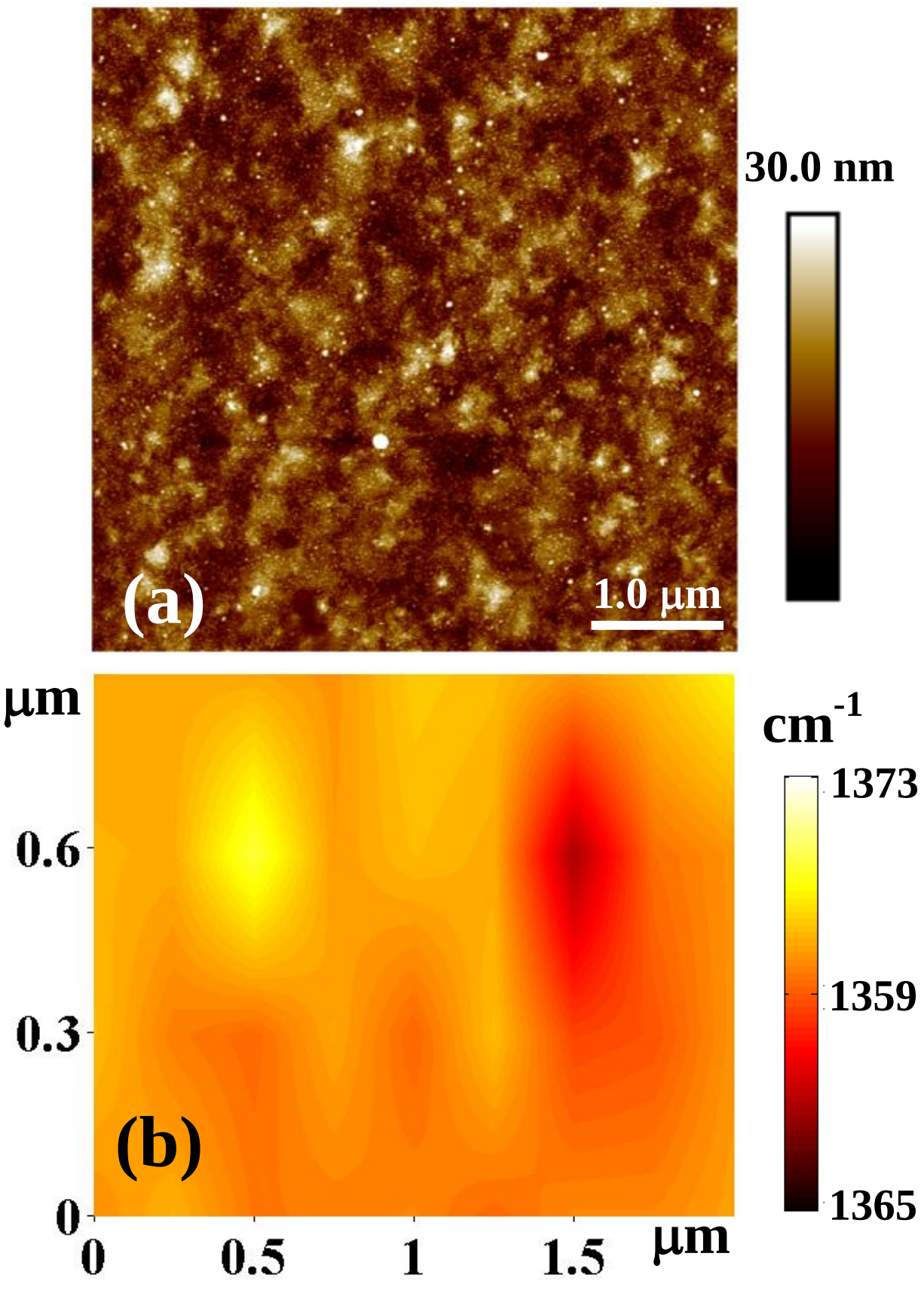

**Figure 6. (a) AFM image of the 20 nm thick film transferred on $SiO_2$/ Si substrate. (b) Mapping of the Raman shift of the $E_{2g}$ mode of the same film. The colour bar in the right of the panel is the scale used for the image.**

## DISCUSSION

### A. The formation of wrinkles

The physics of wrinkling of a film on a substrate is complex and depends on various parameters, like elastic properties, the thickness of both film and substrate [12,13]. The strain distribution in the wrinkled domain is critically governed by the kinetics of the wrinkling. During the growth by MOVPE technique at 1350$^o$C, random pinning of the film with the substrate is expected by weak Lennard-Jones interaction between the two. The formation of wrinkles in the hBN films, shown in Figure 1, can be understood qualitatively in view of the difference in the lattice thermal expansion coefficients of hBN and sapphire. Thin hBN film has a negative in-plane thermal expansion coefficient [27,28], whereas sapphire exhibits the opposite behaviour. After the growth is stopped and the substrate is allowed to cool, while the sapphire tries to contract during lowering of the temperature, the hBN film has a tendency to expand. The pinned boundaries/edges prevent the sheet from expanding laterally. Thus, due to the local biaxial stress, the film is sheared at the edges. Slightly away from the edges, transverse compressive stress develops to nullify the tensile biaxial deformation [12]. The effect is manifested by the bending (local decohesion) of the film from the substrate [12].

### B. Estimation of the residual strain

As mentioned earlier, the origin of strain during the cooling of the MOVPE grown films is mostly governed by the differential thermal expansion coefficients of hBN and sapphire. For similar substrates with positive thermal expansion coefficient, it has been shown that the evolution of the strain in hBN films follows a linear relation with the change is temperature by following the relation [29,30]

$$\varepsilon_{th} = \int_{T_0}^{T_f} (\alpha_s - \alpha_f)\, dT \qquad (1)$$

The above expression is expected to be valid for a small strain in the films [29]. In the present case of the MOVPE grown films, the initial ($T_0$) and final ($T_f$) temperatures are 1623K (1350 $^oC$) and 300K (27 $^oC$), respectively (recall that the strain develops while cooling the films from 1350$^o$C after the growth). $\alpha_f$ and $\alpha_S$ are the linear thermal expansion coefficients of the hBN film and sapphire substrate, respectively. Using, $\alpha_f = -2.96\times10^{-6} + 1.91\times10^{-9}T$ $^oC^{-1}$ [30] and $\alpha_S = 4.5\times10^{-6} + 6.2(T+273)\times10^{-9}T$ $^oC^{-1}$ [31], the calculated biaxial compressive strain ($\varepsilon_{th}$) is typically 1.6%. Wrinkling reduces a part of this compressive strain, however, may not yield a strain free film.

The residual strain ($\varepsilon_r$), left in the films after the formation of the wrinkles, could be revealed from the mapping of the Raman shift of the in-plane $E_{2g}^{high}$ mode of hBN (shown in Figure 3(a)-(d)). We find that the mean value of the Raman wavenumber increases with the increase in layer thickness (Figure 4(a)). For hexagonal 2D layers, Raman shift of the $E_{2g}^{high}$ mode is extremely sensitive to the lattice strain [14,25]. The blue shift in Raman wavenumber indicates an increase in the value of $\varepsilon_r$ with the increase in film thickness.

If we compare the phonon frequencies of the film and bulk hBN, the difference in these two values is expected to carry the information of the residual strain ($\varepsilon_r$) in the films. Due to the hydrostatic and shear strain (to the first order), the shift in phonon frequency in 2D hexagonal layers is given by [14, 32]

$$\Delta\omega = \Delta\omega^h \pm \frac{1}{2}\Delta\omega^s = -\omega_{us}\gamma(\varepsilon_{ll} + \varepsilon_{tt}) \pm \frac{1}{2}\beta\omega_{us}(\varepsilon_{ll} - \varepsilon_{tt}). \quad (2)$$

$\gamma$ is the Grüneisen parameter, which depends on the hBN film thickness. $\beta$ corresponds to the shear deformation potential in the lattice. $\omega_{us}$ is the $E_{2g}^{high}$ band frequency of the unstrained bulk hBN. $\varepsilon_{ll}$ and $\varepsilon_{tt}$ are strain tensor components, $l$ being parallel to the direction of strain and $t$ is perpendicular to the same. Symmetry breaking $(\varepsilon_{ll} \neq \varepsilon_{tt})$ of the lattice can give rise to a splitting of the $E_{2g}$. For example, a uniaxial strain in hBN layer can be characterized by the splitting of this mode [25]. However, under a biaxial strain with $\varepsilon_{ll}=\varepsilon_{tt}$, the $E_{2g}$ mode does not show any splitting due to the cancellation of the shear deformation. As we have not observed any splitting of the $E_{2g}$ mode even when measured at 150K and 78K (not shown), it is reasonable to believe that only a biaxial strain with $\varepsilon_{ll}=\varepsilon_{tt}$ is present in the hBN films. In Eqn. 2, if only the hydrostatic component of the net strain tensor contributes in the phonon frequency of the $E_{2g}^{high}$ mode [14], the corresponding change in Raman shift can be estimated from the relation [14,25]

$$\varepsilon_r = -\Delta\omega / 2\gamma\omega_{us} \qquad (3)$$

A positive value of $\Delta\omega$ (blue shift in Raman wavenumber) renders a compressive strain, while for a tensile strain $\Delta\omega$ is negative.

For different film thickness, we have taken the $\gamma$ values from Ref. [22]. The estimated values of the residual strain, $\varepsilon_r$, of the films of different thicknesses are plotted in Figure 7(a) by green filled up-pointing triangle symbols. We find the higher compressive strain in the thicker films. The expected strain relaxation ($\varepsilon_w$) in the films by wrinkling is then estimated by considering $\varepsilon_w=\varepsilon_{th}-\varepsilon_r$ and plotted in Figure 7(a) by red filled down-pointing triangle symbols. We fitted the data points with empirical relations

$$\varepsilon_w = -1.6h^{-0.02} \qquad \varepsilon_{\mathrm{r}} = -0.02h^{0.5} \qquad (4)$$

The above expressions provide the expected values of $\varepsilon_w$ and $\varepsilon_r$ for a given thickness of the film. As mentioned earlier, the physics of wrinkling is complex. The same for hBN/sapphire system is still unexplored in the literature. For various film-substrate systems, the strain relaxation by wrinkling has been related to the ratio of the wrinkle wavelength ($\lambda$) and thickness ($h$) of the film [12]. Here, by fitting $\varepsilon_w$ in Figure 7(b) with a polynomial function of

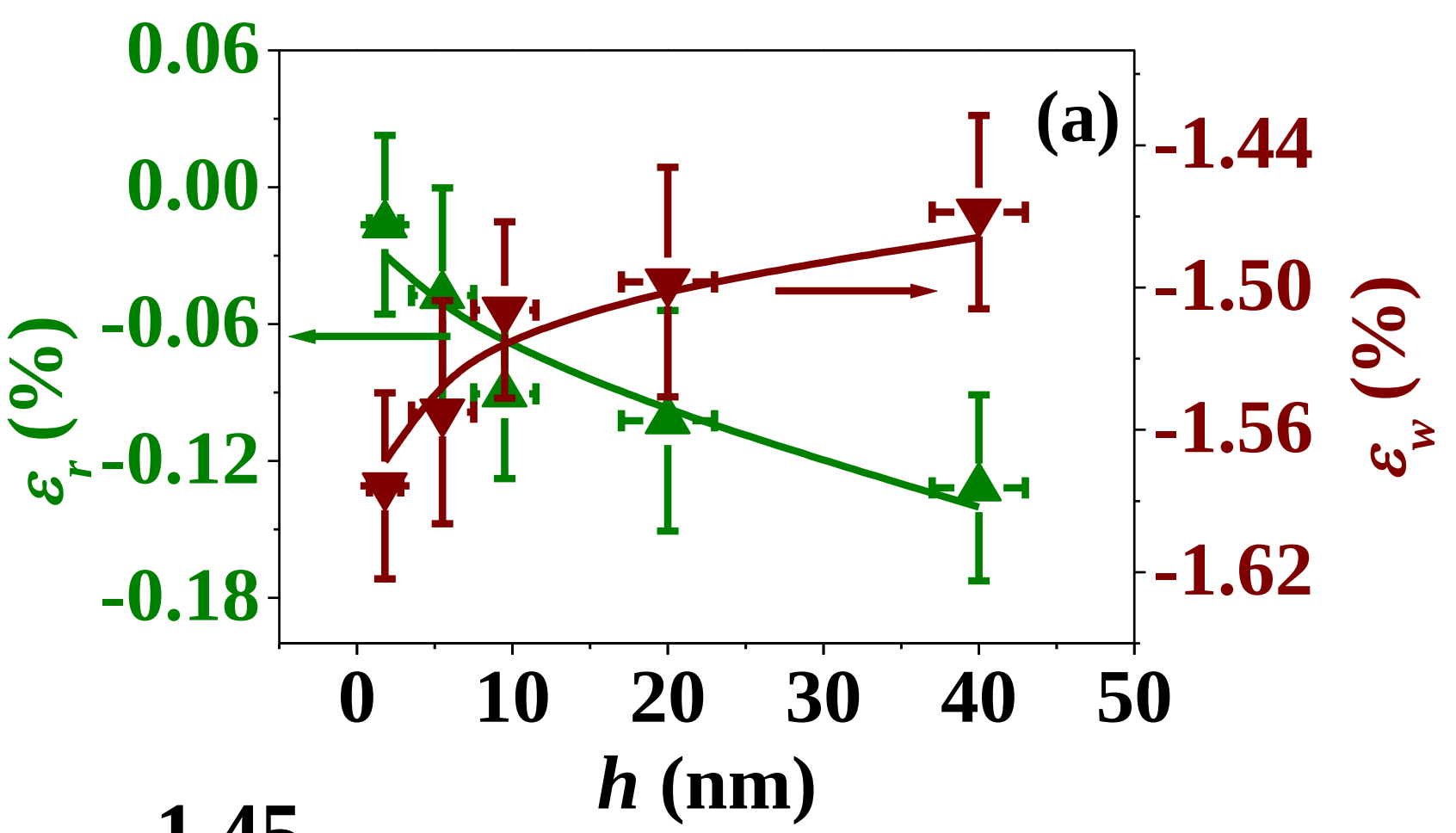

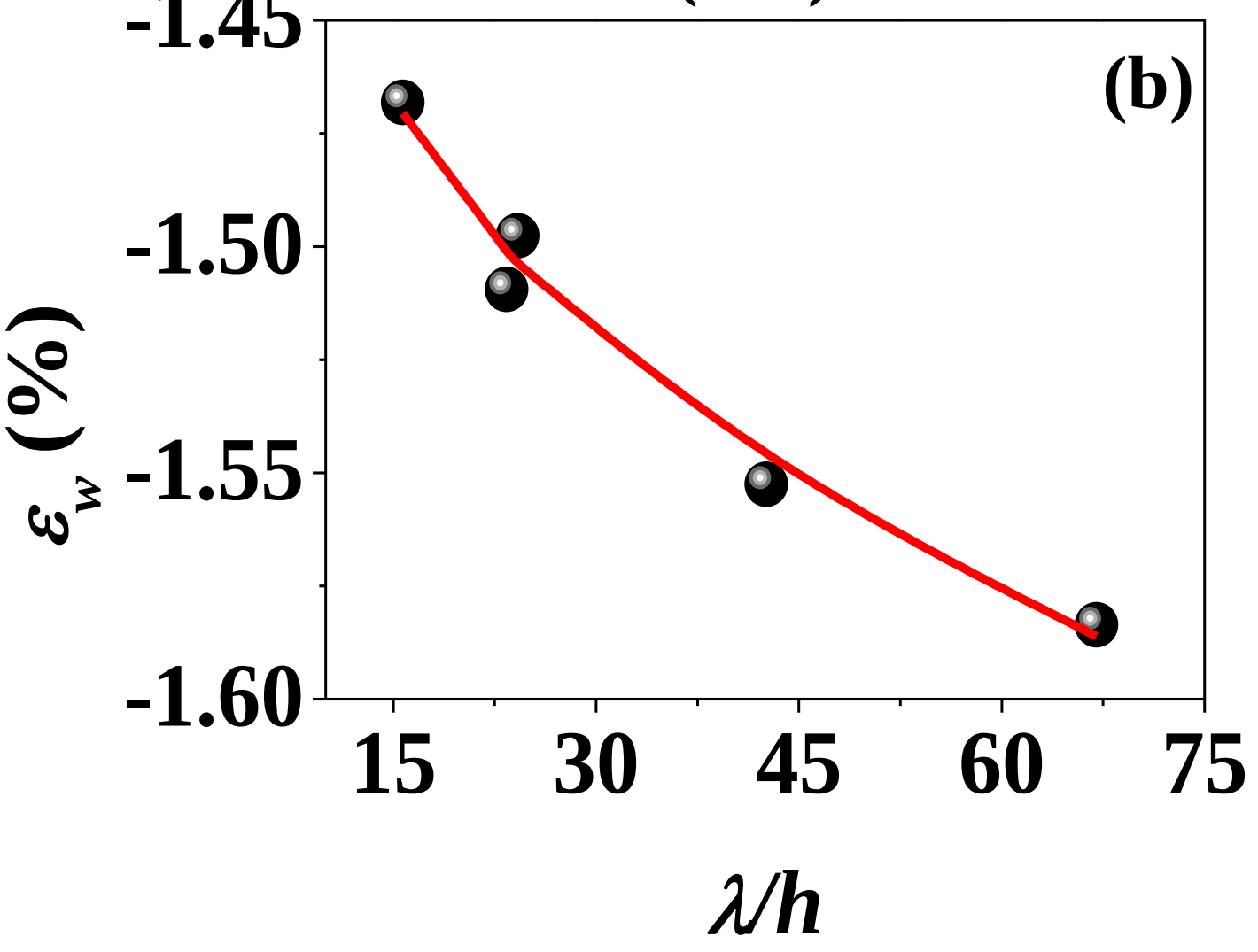

**Figure 7. (a) Variation of residual strain ($\varepsilon_r$) (left scale) and strain relaxation ($\varepsilon_w$) by wrinkling (right scale) with the thickness of the film. The data points are fitted with empirical relations (see text) and shown by the solid lines. (b) Change in $\varepsilon_w$ with $\lambda/h$. The solid line is the fit the data point using the empirical relation, discussed in the text.**

$(\lambda/h)$ we propose an empirical relation

$$\varepsilon_w = -0.3\left(\frac{\lambda}{h}\right)^{0.05} \qquad (5)$$

for an estimation of the strain relaxed by wrinkling from the surface morphology of the hBN films on sapphire.

Raman line width is sensitive to crystallinity and strain in solids. For 2D films, the spectral width is far less sensitive to lattice strain than the Raman shift. For example, in case of graphene, Raman shift (ω) and line width (Γ) vary with strain by dω/dε=60 $cm^{-1}$ and dΓ/dε=12 $cm^{-1}$ [24]. Thus, we believe that the decrease in FWHM of the $E_{2g}^{high}$ mode, observed in Figure 4(b), reflects an improvement in the crystallinity of the films with their thickness.

**C. Thermal response of $E_{2g}^{high}$ mode of hBN/sapphire**

We exploit the temperature dependent Raman measurements, shown in Figure 5, to find the origin of the increase in $\varepsilon_r$ with the increase in the thickness of the film. Further, the same study also probes the possibility of obtaining a strain-free film by thermal treatment. As mentioned earlier, hBN has a negative thermal coefficient, *i.e.*, the lattice parameter is expected to decrease with an increase in the temperature of the film. This gives rise to an increase in phonon frequency ($\omega_{lattice}$) with an increase in the temperature. For the $E_{2g}^{high}$ phonon mode under study, the thermal phonon frequency shift due to lattice contraction of hBN is given by (to the first order approximation) [21]

$$\Delta\omega_{lattuce}(T) = -2.347 + 2.137\times10^{-2}T \quad \text{for } T \geq 160\text{K} \ldots\ldots\ldots\ldots\ldots..(6)$$

Furthermore, the anharmonicity mediated self-energy of phonons, corresponding to three-phonon and four-phonon decay processes, need to be taken into account while studying the variation of phonon frequency with temperature. These processes appear as cubic and quartic terms in the anharmonic Hamiltonian of the oscillator and give rise to a shift in phonon frequency by [33]

$$\Delta\omega_{anh} = A\left(1+\frac{2}{e^{\hbar\omega/2k_BT}-1}\right)+B\left(1+\frac{3}{(e^{\hbar\omega/3k_BT}-1)}+\frac{3}{(e^{\hbar\omega/3k_BT}-1)^2}\right),\ldots\ldots\ldots\ldots(7)$$

where, *A* and *B* are anharmonic constants. In addition, for hBN films, four phonon scattering interaction, another fourth order term in the anharmonic potential, contributes to the phonon self energy. The zone center phonon and two other phonons of opposite wavevectors, participate in this scattering process. These anharmonic effects of light scattering due to phonon gives rise to a decrease in the phonon frequency with an increase in temperature. Using density functional theory calculations, Cusco et al. [21] have shown that the four phonon scattering process introduces substantial variation only for low frequency phonon mode ($E_{2g}^{low}$ at 50 cm$^{-1}$). Thus, to study the change in phonon frequency with temperature of the layered hBN films, shown in Figure 5, we considered the sum contribution of lattice contraction and three and phonon decay processes as

$$\omega(T)= \omega_0+\omega_{lattice}(T)+ \omega_{3pd}(T)+ \omega_{4pd}(T)\text{.........................} (8)$$

$\omega_0$ is the phonon frequency at $T$=0. $\omega_{lattice}(T)$ is the modified phonon wavenumber due to lattice contraction with the increase in temperature by following Eqn. 6. The last two terms in the relation (8) correspond to the Raman shift due to three and four phonon anharmonic decay processes.

First, we fit the $\omega$ vs. $T$ plot of bulk hBN, in Figure 5(a), over the temperature range between 300K and 450K using the Eqn.

$$\omega(T) = \omega_0 + A\left(1+\frac{2}{e^{\hbar\omega/2k_BT}-1}\right)+B\left(1+\frac{3}{(e^{\hbar\omega/3k_BT}-1)}+\frac{3}{(e^{\hbar\omega/3k_BT}-1)^2}\right)+C+DT, \ \dots (9)$$

keeping $C$ and $D$ as $-2.347$ cm$^{-1}$ and $2.137\times10^{-2}$ cm$^{-1}$/K, as in Eqn. 3. $\omega_0$, $A$ and $B$ are kept as free fitting parameters. For the best fit to the data point, we obtained $\omega_0$= 1384 cm$^{-1}$, $A$= –14.35 cm$^{-1}$ and B= –3.27 cm$^{-1}$. While analysing the temperature variation of the Raman shift of the hBN films, under study, we do not expect the anharmonicity in the film to change drastically with a decrease in the layer thickness. However, it may not be exactly the same as obtained for bulk hBN. The parameter $D$, related to change in Raman shift with temperature due negative thermal expansion coefficient, also can be modified with the thickness of the film. Furthermore, while rising the temperature during the temperature dependent Raman measurements, due to differential thermal expansion coefficient of the films and the substrate, additional strain is expected to develop in the films. Assuming that the strain varies linearly with temperature (as in Eqn. 1) for the given film-substrate system, the parameter $D$ can be considered as the thermal strain coefficient of Raman wavenumber of the films.

Using above arguments and Eqn. 8 we fitted the data points in all panels in Figure 5 by

$$\omega(T) = \omega_0 - 14.45\left(1 + \frac{2}{(e^{\frac{\hbar\omega_0}{2k_BT}} - 1)}\right) + B\left(1 + \frac{3}{(e^{\frac{\hbar\omega_0}{3k_BT}} - 1)} + \frac{3}{(e^{\frac{\hbar\omega_0}{3k_BT}} - 1)^2}\right) - 2.347 + DT \qquad (10)$$

$\omega_0$, $B$ and $D$ are kept as free fitting parameters. The variation in $B$ is expected to take care of the change in phonon frequency arising due to change in anharmonicity with the thickness of the film. The change in the parameters $B$ and $D$ with the change in the thickness of the film are plotted in Figure 8. The error bars to the data points are the standard deviation of the parameters, obtained from the fitting procedure. In Figure 8(a) we find that the anharmonicity does not change appreciably with the film thickness and the values of the anharmonic coefficient for all films are very close to the same obtained for bulk hBN. Interestingly, we find a large change in the value of the parameter *D*, as shown in Figure 8(b). In fact, while the value of *D* is positive for bulk hBN, it clearly has a large negative value for the 2 nm film. Such a large change is unlikely due to the variation in lattice thermal expansion coefficient with the thickness of the film. It is to be noted that the lattice thermal expansion coefficient depends on anharmonicity in vibrational potential in a solid. Keeping in mind that the anharmonic coefficient *A* was kept fixed and the variation in *B* is negligible with the change in thickness of the film (as shown in Figure 8(a)), we believe that the parameter *D* dominantly reflects the thermal strain coefficient of Raman wavenumber in the film on the sapphire substrate. The relatively higher value of *D* for the thinner films than for the films of thickness > 10 nm, implies that the change in Raman shift with temperature is higher in case of thinner films than in thicker films.

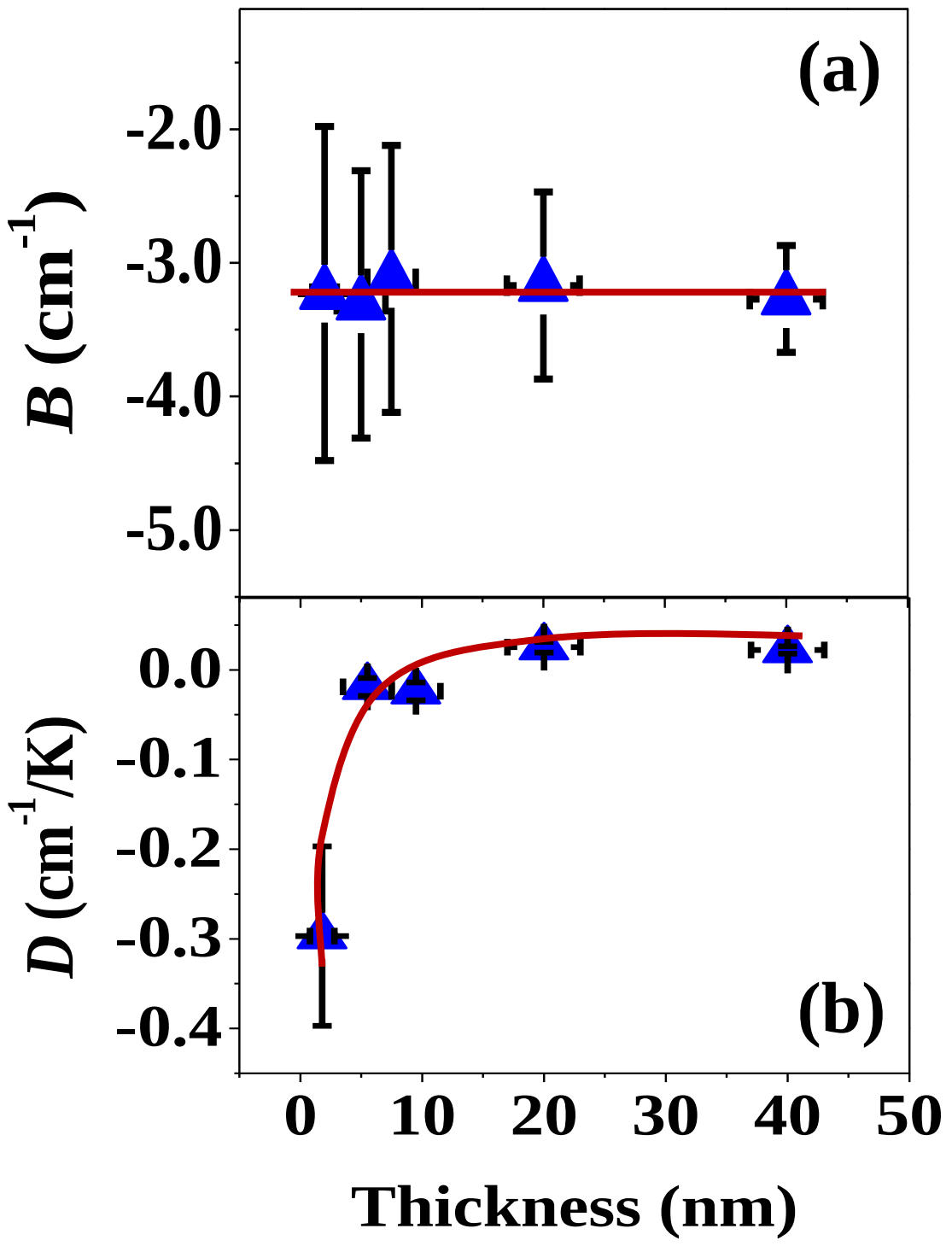

**Figure 8. Variation of the parameters (a) *B* and (b) *D* with the thickness of the films.**

We exploit above finding to understand the increase in compressive residual strain (*i.e.* Raman shift) with the film thickness, shown in Figure 4. It is to be recalled that the compressive strain develops due to the differential thermal expansion coefficient between the hBN layer and the substrate, while cooling the MOVPE grown films from 1300$^{o}$C to room temperature. Due to the higher value of *D* for the thinner than the thicker films, one would expect a higher value of the residual strain, and hence the Raman shift, in case of the former at the room temperature. This is opposite to what we find in Fig. 4(a). We believe that the higher strain relaxation by wrinkling reduces the compressive strain in the films of lower thickness. Indeed, this is what could be revealed from our analysis in Figure 7(a).

We have obtained a systematic change in $\omega_0$ with film thickness. However, the value of $\omega_0$ does not have much significance in the present analysis, as the thermal expansion coefficient of hBN is different from Eqn. 6 below 160K [21].

**(D) Reduction in strain by thermal treatment and delamination**

We find from our earlier discussion that the wrinkling leaves a non-uniform Raman shift, and hence a non-uniform strain distribution over the surface of the film. In addition, we carry out high temperature Raman mapping of the films. Exploiting the imaging of the Raman shift, Figure 9(a) and (b) plot the mapping of the residual strain, $\varepsilon_r$ in the 2 nm film at room temperature and at 323 K, using Eqn. 3 . In this analysis we have used the $\omega_{us}$ values for bulk hBN at 300K and 323K as 1367, 1366.5 $cm^{-1}$, respectively. Here we use the above arguments that the effect of phonon anharmonicity and the thermal expansion coefficient on the phonon frequency are same both in the film and bulk. It is interesting to note that at 323 K the film shows a tensile strain. This observation further indicates that the rise in temperature results in an additional strain in the film, as we conjectured above. One expects a tensile strain with increase in temperature (recall that the lowering of temperature in the MOVPE chamber resulted a compressive strain). Thus, it is reasonable to conclude that at a certain temperature this tensile strain can compensate the residual compressive strain to yield a strain-free film.

To investigate whether it is possible to obtain strain-free films of various thicknesses, we use the fitted plots of temperature variation of Raman shift of bulk hBN and films, presented in Figure 5. In Figure 9(c) the magenta, blue, red, green and cyan solid lines are the fitted solid lines from Figure 5, showing the evolution of Raman shift with temperature of 40 nm, 20 nm, 10 nm, 5 nm and 2 nm films respectively. The black solid line is the same for bulk hBN. The intercept of the fitted curve for a film and bulk hBN provides the temperature required to nullify the residual strain in the former. As mentioned earlier, due to the standard deviation of the Raman shift over the wrinkled domains (shown in Fig. 4(a)), the fitted curve in Figure 5 may upshift or downshift. The intercept discussed above, changes for a given initial Raman shift. For example, the magenta dashed lines in Fig. 9(c) mark the boundaries of the Raman shift vs. temperature plots for 40 nm and the red arrow (bottom right) marks the corresponding temperature range for the intercepts. The vertical bar for each film thickness in Fig. 9(d) encompasses this range of temperature. Importantly, the error bars to the data points shown in Fig. 5 is much less than the spread in Raman wavenumber due non-uniform surface morphology of the films.

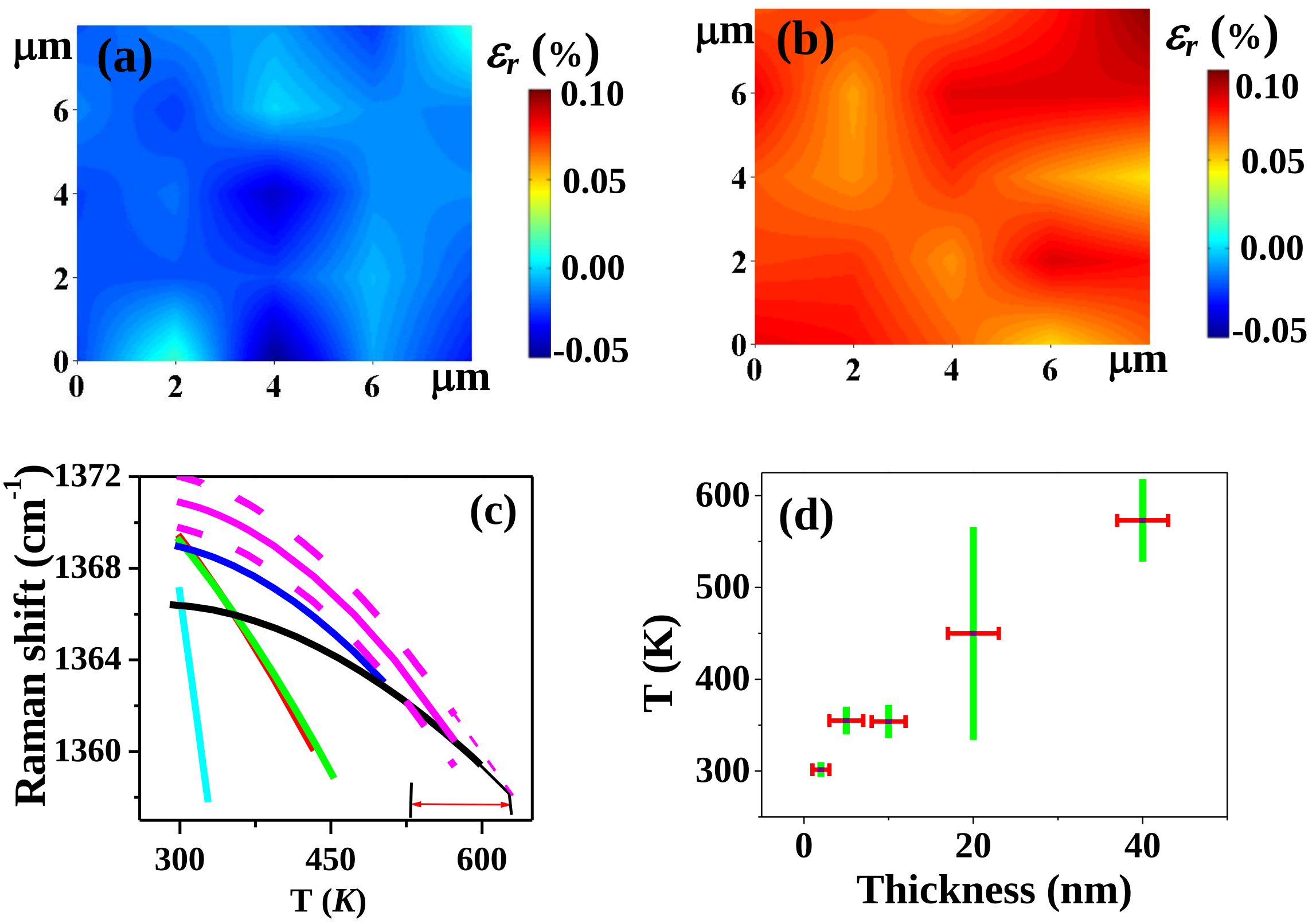

**Figure 9. Mapping of the residual strain for the 2 nm film at (a) 300K and (b) 323K. (c) The analysis to obtain the null strain (see the text). Black, magenta, blue, red, green and cyan solid lines represent temperature dependent Raman shift, the fitted line from Figure 5 for bulk hBN and films of thickness 40 nm, 20 nm, 10 nm, 5 nm and 2 nm, respectively. The dashed magenta lines are plots to show the possible variation of Raman shift vs. temperature for 40 nm film. The red arrow in the bottom right shows the temperature range over which the magenta dashed lines for 40 nm film intercept the plot for the bulk. (d) Estimated temperature range to obtain nearly strain-free hBN films of different thicknesses.**

Though, for the films of higher thickness, the spread in temperature is quite high, for the thinner layers (<10 nm) the result is promising. For these films, it may not be possible to obtain a fully strain-relaxed film by thermal treatment over a wide area. However, the study provides a narrow range of temperature to scan to obtain a nearly strain-free film. However, locally the small residual strain may remain in the layers.

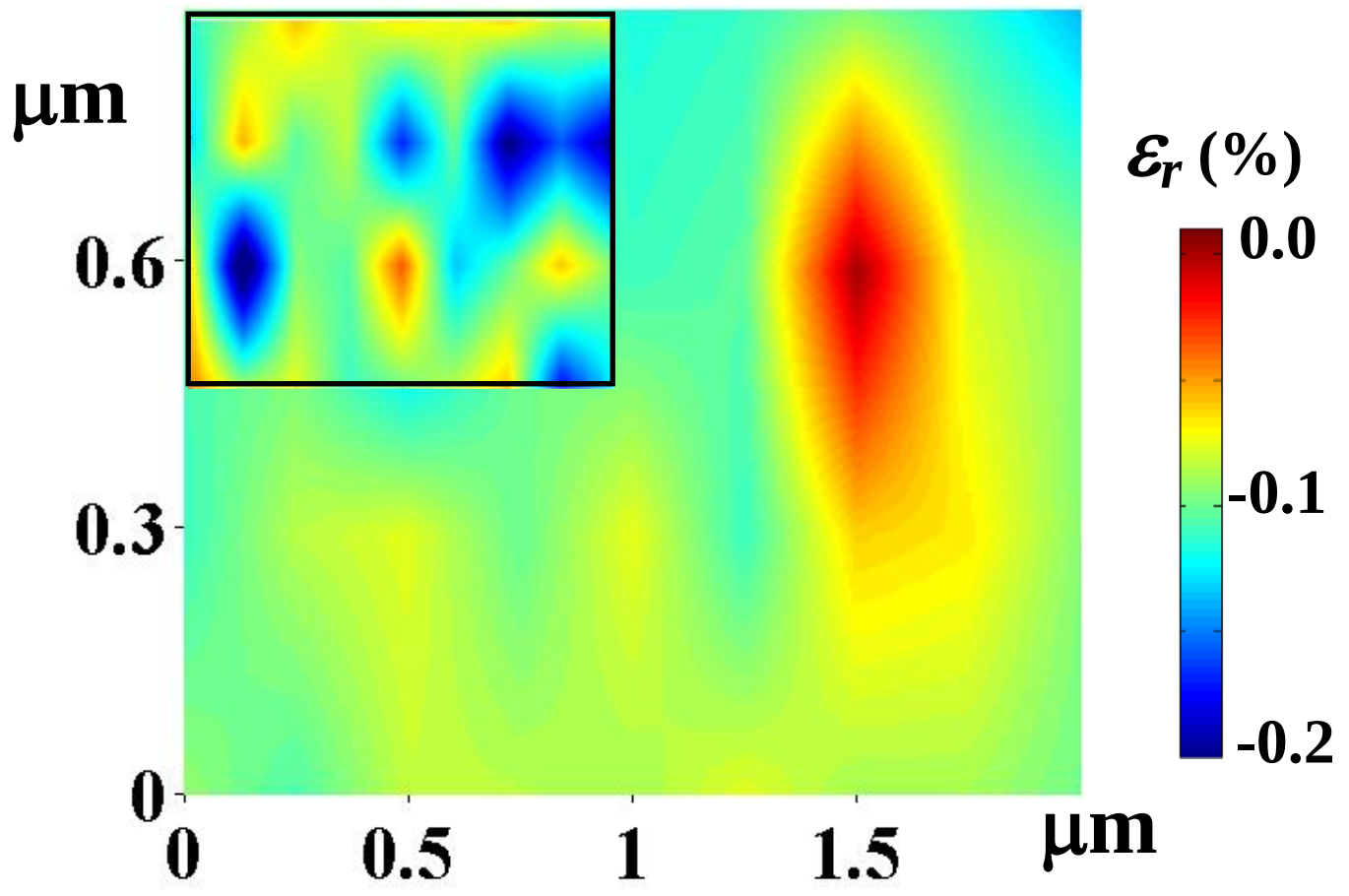

**Figure 10. Mapping of the residual strain for the 20 nm delaminated film. The same for the substrate supported film is shown in the inset using the same scale bar.**

It is to be noted that the plot renders an easy measure to find the temperature to obtain a nearly strain-free hBN film of a given thickness on sapphire substrate.

Here we would like to mention that to probe the reversibility of the strain relaxation process, we first raised the temperature of the films and then lowered it to room temperature. We obtained the spectral profile similar to what observed in the beginning of the cycle. Thus, though the strain in these films can be modulated by thermal treatment, the process is reversible. The films attains its original strain profile once we bring it back to the room temperature

In Figure 6(a), the AFM image of the transferred 20 nm film indicates that delamination reduces the wrinkling. In Figure 6 (b) we have shown the mapping of Raman shift of the $E_{2g}^{high}$ mode for the transferred film of thickness 20 nm. Comparing the same in Figure 3(c) for sapphire-supported film we find a lowering of the mean Raman shift and its narrower distribution in the transferred film, indicating a strain relaxation upon delamination. The strain mapping of the 20 nm delaminated film is shown in Figure 10, The same of the substrate-supported film is shown in the inset. We find that the strain distribution is significantly low in the delaminated film.

**Summary**

Thin films of hBN on sapphire substrates are grown by FM-MOVPE technique. While cooling the samples in the reaction chamber, wrinkles develop on the films. The morphology of the films is characterized by AFM images. The wrinkles leave a non-uniform residual strain distribution in the films, which has been probed by Raman mapping. We have observed an overall compressive strain in the films. The variation of strain distribution with the thickness of the films is discussed. Empirical relations are proposed for the estimation of residual strain and strain relaxed by wrinkling from the morphology of the films. Furthermore, we have carried out temperature dependent Raman measurements of all films under study, to (i) investigate whether the residual strain can be removed by thermal treatment, (ii) find the origin of the variation in the residual strain with the thickness of the films and (iii) propose the temperature at which one can obtain strain-free substrate-supported films.

The presence of strain in a film may not be desirable. The performance of hBN film based devices can get affected due to the intrinsic strain in the system. Thus, we believe that the present study not only discusses the strain profile of the as-grown hBN films of varying thicknesses, but also reveals possible ways to obtain strain-free hBN films for future applications.